\documentclass[journal, 11pt, onecolumn, draftclsnofoot]{IEEEtran}
\usepackage[usenames, dvipsnames]{color}
\usepackage[left=1in, right=1in, top=1.25in, bottom=1.25in]{geometry}
\usepackage{citesort}
\usepackage{setspace}
\usepackage{pifont}
\ifCLASSINFOpdf
  \usepackage[pdftex]{graphicx}
	\usepackage{epstopdf}
\else
  \usepackage[dvips]{graphicx}
\fi
\usepackage{moresize}
\usepackage[cmex10]{amsmath}
\usepackage{algorithmic}
\usepackage{amssymb}
\usepackage{array}
\usepackage{mdwmath}
\usepackage{mdwtab}
\usepackage{eqparbox}
\usepackage{multirow}
\usepackage{fixltx2e}
\usepackage{enumerate}
\usepackage{url}
\usepackage{tabularx}
\usepackage{verbatim}
\usepackage{color}
\usepackage{soul}

\usepackage{stmaryrd}
\usepackage{bm}
\usepackage{amssymb}
\usepackage{booktabs}
\usepackage{graphicx}
\usepackage{bbding}
\usepackage{float}
\usepackage{graphicx}
\usepackage{stfloats}
\usepackage{amsmath}
\usepackage{epstopdf}
\usepackage{algorithm,algorithmic}
\usepackage{multirow}
\usepackage{bbm}
\usepackage{framed}
\usepackage[colorlinks,linkcolor=red,anchorcolor=green,citecolor=green]{hyperref}
\usepackage[numbers,sort&compress]{natbib}
\usepackage{amsthm}

\DeclareMathOperator*{\argmin}{arg\,min}

\newcommand{\ie}{{\em i.e.}}
\newcommand{\eg}{{\em e.g.}}

\renewcommand{\IEEEbibitemsep}{0pt plus 0.5pt}
\makeatletter
\IEEEtriggercmd{\reset@font\normalfont\fontsize{7.7pt}{8.20pt}\selectfont}
\makeatother
\IEEEtriggeratref{1}

\begin{document}
\title{Learning Nonlocal Sparse and Low-Rank Models for Image Compressive Sensing}
\author{Zhiyuan~Zha,~\IEEEmembership{Member,~IEEE}, Bihan~Wen,~\IEEEmembership{Member,~IEEE}, Xin~Yuan,~\IEEEmembership{Senior Member,~IEEE},\\ Saiprasad~Ravishankar,~\IEEEmembership{Senior Member,~IEEE}, Jiantao~Zhou,~\IEEEmembership{Senior Member,~IEEE} and  Ce~Zhu,~\IEEEmembership{Fellow,~IEEE}
\IEEEcompsocitemizethanks{
\IEEEcompsocthanksitem Zhiyuan Zha and Bihan Wen are with the School of Electrical \& Electronic Engineering, Nanyang Technological University, Singapore 639798.  (E-mail: zhiyuan.zha@ntu.edu.sg; bihan.wen@ntu.edu.sg.)
\IEEEcompsocthanksitem Xin Yuan is with School of Engineering, Westlake University, Hangzhou, Zhejiang 310024, China. E-mail: xyuan@westlake.edu.cn.
\IEEEcompsocthanksitem Saiprasad Ravishankar is with the Department of Computational Mathematics, Science and Engineering, and the Department of Biomedical Engineering, Michigan State University, East Lansing, MI, 48824 USA. Email: ravisha3@msu.edu.
\IEEEcompsocthanksitem Jiantao Zhou is with the State Key Laboratory of Internet of Things for Smart City, and Department of Computer and Information Science, University of Macau, Macau 999078, China. E-mail: jtzhou@um.edu.mo.
\IEEEcompsocthanksitem Ce Zhu is with the School of Information and Communication Engineering,
University of Electronic Science and Technology of China, Chengdu, 611731, China.  E-mail:  eczhu@uestc.edu.cn.
}
}
\maketitle

\vspace{-0.6in}
\begin{abstract}
The compressive sensing (CS) scheme exploits much fewer measurements than suggested by the Nyquist-Shannon sampling theorem to accurately reconstruct images, which has attracted considerable attention in the computational imaging community. While classic image CS schemes employ sparsity using analytical transforms or bases, the learning-based approaches have become increasingly popular in recent years. Such methods can effectively model the structure of image patches by optimizing their sparse representations or learning deep neural networks, while preserving the known or modeled sensing process. Beyond exploiting local image properties, advanced CS schemes adopt nonlocal image modeling, by extracting similar or highly correlated patches at different locations of an image to form a group to process jointly. More recent learning-based CS schemes apply nonlocal structured sparsity priors using group sparse (and related) representation (GSR) and/or low-rank (LR) modeling, which have demonstrated promising performance in various computational imaging and image processing applications. This article reviews some recent works in image CS tasks with a focus on the advanced GSR and LR based methods. Furthermore, we present a unified framework for incorporating various GSR and LR models and discuss the relationship between GSR and LR models. Finally, we discuss the open problems and future directions in the field.
\end{abstract}

\begin{IEEEkeywords}
\vspace{-0.1in}
Compressive sensing, sparse representation, computational imaging, nonlocal self-similarity, group sparsity, low-rank, dictionary learning, transform learning, inverse problems, optimization.
\end{IEEEkeywords}

\vspace{-0.15in}
\section{Introduction}

\IEEEPARstart{I}{n} this era of information explosion, billions of images and videos are rapidly spread by social networks and mobile Internet. Therefore, an efficient sensing and compression scheme is extremely critical for managing these large-scale multimedia data that are kept in expensive storage and transmission devices~\cite{Candes2006Robust,Candes2008An}. Traditional data (signal) acquisition and compression schemes follow the classic Nyquist-Shannon sampling theorem to remove signal redundancy and maintain the essential information of signals, \ie, the sampling rate is no less than twice of the signal bandwidth. More recent compressive sensing (CS) theory \cite{Candes2008An} showed that a signal can be accurately recovered from much fewer measurements than suggested by the Nyquist-Shannon sampling theorem, which indicates great potential for simultaneously improving the sensor, data storage and transmission efficiency in practical applications. One of the most well-known successes of CS is in magnetic resonance imaging (CS-MRI) \cite{Ravishankar2011MR}, where CS helps significantly reduce the number of samples and the acquisition time in MRI scanning, while maintaining the high-quality reconstructions. In particular, the Food and Drug Administrations (FDA) of many countries have approved such CS-MRI techniques for clinical use, paving its use for assisting doctors in practice. More CS-based imaging systems have been established, including the single-pixel camera \cite{Duarte2008Single}, high-speed video camera \cite{Hitomi2011Video}, and compressive spectral imaging system \cite{Gehm2007Imaging}, providing huge commercial values to the industry.

\begin{figure*}[!t]
\vspace{-2mm}
\centering
\begin{minipage}[b]{1\linewidth}
\centering
{\includegraphics[width = 0.60\textwidth]{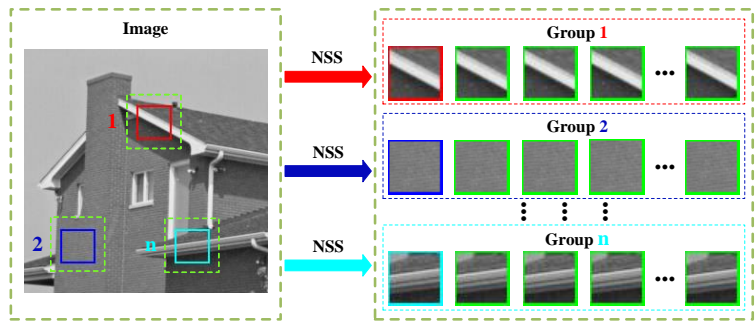}}
\end{minipage}
\vspace{-8mm}
\caption{Illustration of image NSS prior. Taking the small red square as an exemplar patch within a big green dashed square, and then a variant of the $K$-nearest-neighbor (KNN) algorithm is conducted to search its similar patches (small green square) to form a group.}
\label{fig:1}
\vspace{-8mm}
\end{figure*}

While early image CS methods employed sparsity using analytical transforms (\eg, wavelet \cite{Dong2014Sparsity} and total variation \cite{li2009user}), recent years have witnessed an increasing interest in learning the underlying image models for reconstruction \cite{Zhang2014Image,Dong2014Compressive,Zha2020Group,Wang2021Hybrid,Kulkarni2016ReconNet}. The learning-based CS methods integrate model-based learning (\eg, sparse representation \cite{Zha2020Group} or deep learning \cite{Shi2020Image}) to achieve superior results in imaging tasks, such as dictionary learning-based MRI reconstruction \cite{Ravishankar2011MR}. Such approaches can effectively model the structure of images or image patches by optimizing their sparse representations or learning the deep neural networks \cite{Krizhevsky_NIPS2012,Hong2021Graph}, while preserving the known or modeled sensing process. However, most existing learning-based approaches have mainly considered local image  properties (\eg, exploiting every single image patch as the basic unit for sparse representation \cite{Zhang2014Image,Ravishankar2011MR}), while they have largely ignored image nonlocal self-similarity (NSS) properties \cite{Buades2005A,Dabov2007Image}, as shown in Fig.~\ref{fig:1}. Therefore, advanced CS strategies with nonlocal image modeling (involving grouping similar or highly correlated patches from different locations of an image) can be used for more effective model learning. More recent learning-based CS schemes exploited nonlocal structured sparsity priors using group sparse (and related) representation (GSR) \cite{Zhang2014Group} and/or low-rank (LR) \cite{Dong2014Compressive} modeling, demonstrating promising performance in various computational imaging applications.

In this review article, we first discuss using local image priors for image CS, and then discuss the recent advances, particularly involving NSS prior-based learning methods for image CS. We mainly focus on reviewing the learning of nonlocal structured sparsity models for image CS under a single umbrella, including GSR and LR representations, which produce state-of-the-art reconstruction results. We review different types of GSR and LR learning schemes with their model structures and properties, and discuss how these nonlocal learning methods can be effectively employed in image CS under a unified framework. We also consider the relationship between GSR and LR models, and discuss the open  problems and future directions in the field. The goal of this article is not to provide a comprehensive review of all classes of nonlocal modeling-based image CS approaches but rather to focus on the recent GSR and LR learning techniques and discuss their properties, underlying models, benefits, connections, and extensions. At the same time, this article mainly provides the readers with methodologies for image CS.

\vspace{-0.15in}
\section{Image CS: From Local to Nonlocal Methods}
\label{sec:2}

\begin{figure*}[!t]
\vspace{-2mm}
\centering
\begin{minipage}[b]{1\linewidth}
{\includegraphics[width = 1\textwidth]{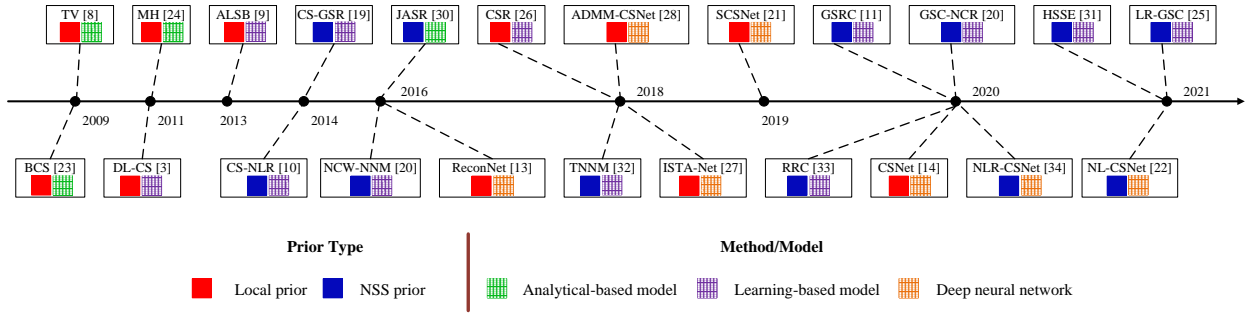}}
\end{minipage}
\vspace{-8mm}
\centering
\caption{A timeline illustrating the evolution of classic local prior-based image CS to recent nonlocal prior and learning-based image CS approaches. Only a few papers are included as examples of each class of approaches (categories are not strictly chronological).}
\label{fig:2}
\vspace{-8mm}
\end{figure*}

Image CS exploits random sampling technologies that extract a few measurements and use them to reconstruct the original image. Since the dimensionality of acquired random measurements is far lower than the original image,  image CS is essentially an ill-posed or under-determined inverse problem \cite{Zhang2014Group,Zha2020Benchmark,Shi2020Image}. To tackle this problem, one approach is to use the underlying image priors (\eg, sparsity \cite{Zhang2014Image}, deep image prior \cite{Kulkarni2016ReconNet,Shi2019Scalable} and NSS \cite{Zha2020Group}) to design effective regularizers, which can play a critical role in ensuring accurate image reconstruction. In the past decade, numerous prior-based models have been proposed for image CS, including mainly two categories: local image prior-based models \cite{li2009user,Ravishankar2011MR,Shi2020Image} and nonlocal image prior-based models \cite{Zhang2014Group,Dong2014Compressive,Cui2021Image}. In this section, we review some of the approaches in image CS (reconstruction from limited data) based on local image priors, followed by recent advances in nonlocal image CS reconstruction. Fig.~\ref{fig:2} shows a timeline for the evolution of image CS with local priors to nonlocal image CS in the past years, with some representative works in each class.

\vspace{-0.15in}
\subsection{Local Priors for Image CS}
\label{sec:2.1}

\begin{figure*}[!t]
\vspace{-2mm}
\centering
\begin{minipage}[b]{0.60\linewidth}
{\includegraphics[width = 1\textwidth]{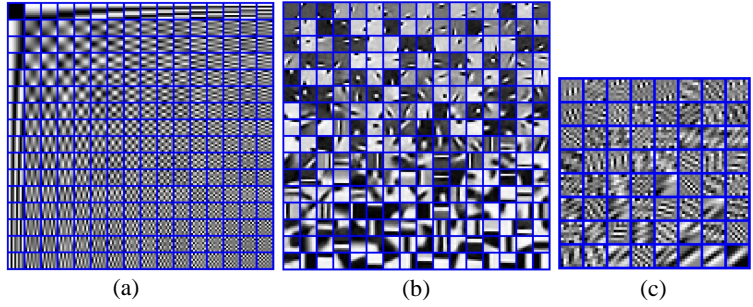}}
\end{minipage}
\vspace{-4mm}
\caption{Illustration showing examples of three dictionaries: (a) fixed DCT dictionary, (b) over-complete learned dictionary, and (c) adaptive principal component analysis (PCA) sub-dictionary.}
\label{fig:3}
\vspace{-8mm}
\end{figure*}

Early image CS approaches assumed that natural images are sparse in certain analytical transform domains, such as wavelet \cite{Dong2014Sparsity} and finite difference (total variation \cite{li2009user}) domains. However, since natural signals (\eg, images) are usually non-stationary, it is hard to find a common sparsifying model that works for all types of images to achieve high-quality reconstruction \cite{Zhang2014Image}. Later works integrated more sophisticated models into the image reconstruction framework, such as patch sparsity prior \cite{Zhang2014Image} and deep image prior \cite{Kulkarni2016ReconNet}. The patch sparsity model usually assumes that each patch of an image can be  modeled as a sparse vector via
analytically designed dictionaries (\eg, DCT/wavelet dictionary). Examples include directional transform-based block CS (BCS) \cite{Mun2009Block} and multi-hypothesis prediction (MH) that generates a residual in the transform domain \cite{Chen2011Compressed}. While such methods use fixed or known transformations (see Fig.~\ref{fig:3}(a) for an example) to achieve fast reconstructions, learning-based approaches have become increasingly popular in recent years, which may learn an image model from a corpus of data and use it during reconstruction time or even learn the model while performing reconstruction from limited measurements (\ie, blind CS \cite{Ravishankar2011MR,Dong2014Compressive,Zha2021Reconciliation}). Typically, the learned sparsity model allows representing each patch of an image with a sparse vector (many of whose elements are zero), wherein the underlying dictionary is learned from a natural image dataset (see Fig.~\ref{fig:3}(b) for an example). Classic learning-based image CS approaches include synthesis dictionary learning as a data-driven regularizer for image CS (DL-CS) \cite{Ravishankar2011MR}, adaptively learned sparsifying basis (ALSB) \cite{Zhang2014Image}, and centralized sparse regularization (CSR) \cite{zha2018compressed}, which can highly adapt to image local structures and are thus more flexible than analytically designed dictionary-based methods \cite{Mun2009Block,Chen2011Compressed}. More recent deep learning-based models have proven effective for image CS by learning local image properties from training datasets with an end-to-end approach. Examples include reconstructed convolutional neural network (ReconNet) \cite{Kulkarni2016ReconNet}, iterative shrinkage threshold algorithm network (ISTA-Net) \cite{Zhang2018ISTANet}, alternating direction method of multipliers-CS neural network (ADMM-CSNet) \cite{Yang2020ADMMCSNet}, scalable convolutional neural network (SCSNet) \cite{Shi2019Scalable}, and CS joint use sampling network and reconstruction network (CSNet) \cite{Shi2020Image}.

\vspace{-0.15in}
\subsection{Nonlocal Image CS}
\label{sec:2.2}
In the past decade, one of the most popular image priors is the NSS, which captures the repetitiveness of textures and structures across nonlocal regions of an image, implying that many similar patches can be sought for each exemplar patch (see Fig.~\ref{fig:1}). Popularized by pioneering image denoising works such as nonlocal-means (NLM) \cite{Buades2005A} and block-matching and 3D filtering (BM3D) \cite{Dabov2007Image}, dozens of methods have been developed that have demonstrated high effectiveness in various image restoration tasks \cite{Gu2014Weighted,Zha2020Group}. In the field of computational imaging, nonlocal image modeling (for CS) extracts similar or highly correlated patches at different locations of an image to construct a group, which can be employed for more effective model learning, resulting in superior performance improvements over local image 
models, such as sparsity fine-tuning of image CS \cite{Dong2014Sparsity} and joint adaptive sparsity regularization (JASR) \cite{Eslahi2016Compressive}. Using image NSS prior and clustering similar patches into a group (see Fig.~\ref{fig:1}), and learning nonlocal structured sparsity priors using GSR and/or LR representations has shown great potential in various image CS studies. Nonlocal structured sparsity prior using GSR and/or LR modeling offers numerous advantages, such as closed-form updates in the iterative algorithms, good convergence, ease in merging a variety of model properties, and effectiveness in reconstruction. Therein, several remarkable GSR-based image CS schemes have been proposed, including group sparse representation for image CS (CS-GSR) \cite{Zhang2014Group} by learning a self-adaptive dictionary; group sparsity residual constraint (GSRC) \cite{Zha2020Group}, which learns an effective group sparsity residual prior with a self-supervised learning scheme; group sparse coding with non-convex regularization (GSC-NCR) \cite{Zha2020Benchmark} that learns a non-convex structured sparsity prior; the hybrid structural sparsification error (HSSE) model \cite{Zha2021Hybrid}  jointly learns group sparsity prior using internal and external image data that provide complementary information for reconstruction. Meanwhile,  LR-based reconstruction schemes have also demonstrated excellent performance in image CS, including CS via nonlocal low-rank regularization (CS-NLR) \cite{Dong2014Compressive}, which estimates the rank of each group matrix by the weighted nuclear norm minimization (WNNM) algorithm \cite{Gu2014Weighted}; non-convex weighted nuclear norm minimization (NCW-NNM) \cite{Zha2020Benchmark}  estimates the rank of each group matrix by solving a non-convex optimization problem; the truncated nuclear norm minimization (TNNM) model  \cite{geng2018truncated}   estimates the rank of each group matrix by truncating the singular values; and the rank residual constraint (RRC) model \cite{Zha2020From}  learns a rank residual to estimate the rank of each group matrix. Beyond employing different NSS priors by learning, the low-rank regularized group sparse coding (LR-GSC) model \cite{Zha2021Reconciliation} integrates different GSR and LR learning formulations into a unified CS framework by reconciling nonlocal image modeling. {The most recent trend combined deep neural networks with NSS prior for image CS, such as learning nonlocal regularized compressed sensing network (NLR-CSNet) \cite{Sun2020Learning}, and image CS framework using nonlocal neural network (NL-CSNet) \cite{Cui2021Image}}. Such nonlocal model-based supervised deep learning image CS schemes have also yielded promising performance. In this article, we focus on learning nonlocal sparse and LR representation methods, which provide flexibility and numerous modeling, computational, convergence, and performance advantages.

\begin{table*}[!t]
\vspace{-2mm}
\caption{A comparison of several types of image CS approaches surveyed in this article.}
\vspace{-4mm}
\centering
\resizebox{1\textwidth}{!}				
{
\scriptsize
\centering
\begin{tabular}{|c|cccc|cccc|cccc|}
\hline
\centering
\multirow{3}{*}{\textbf{Method}} & \multicolumn{4}{c|}{\textbf{Prior Type}}& \multicolumn{4}{c|}{\textbf{Model}}& \multicolumn{4}{c|}{\textbf{Properties}}\\
\cline{2-13}
&\multicolumn{1}{c|}{\multirow{2}{*}{\begin{tabular}[c]{@{}c@{}}\textbf{Internal}\\ \textbf{prior}\end{tabular}}}
&\multicolumn{1}{c|}{\multirow{2}{*}{\begin{tabular}[c]{@{}c@{}}\textbf{External}\\ \textbf{prior}\end{tabular}}}
&\multicolumn{1}{c|}{\multirow{2}{*}{\begin{tabular}[c]{@{}c@{}}\textbf{Local}\\ \textbf{prior}\end{tabular}}}
&\multirow{2}{*}{\begin{tabular}[c]{@{}c@{}}\textbf{NSS}\\ \textbf{prior}\end{tabular}}
&\multicolumn{1}{c|}{\multirow{2}{*}{\begin{tabular}[c]{@{}c@{}}\textbf{Analytical}\\ \textbf{model}\end{tabular}}}
&\multicolumn{1}{c|}{\multirow{2}{*}{\begin{tabular}[c]{@{}c@{}}\textbf{Learning}\\ \textbf{model}\end{tabular}}}
&\multicolumn{1}{c|}{\multirow{2}{*}{\begin{tabular}[c]{@{}c@{}}\textbf{GSR}\\ \textbf{model}\end{tabular}}}
& \multirow{2}{*}{\begin{tabular}[c]{@{}c@{}}\textbf{LR}\\ \textbf{model}\end{tabular}}
& \multicolumn{1}{c|}{\multirow{2}{*}{\begin{tabular}[c]{@{}c@{}}\textbf{Self-}\\ \textbf{supervised}\end{tabular}}}
& \multicolumn{1}{c|}{\multirow{2}{*}{\textbf{Supervised}}}
& \multicolumn{1}{c|}{\multirow{2}{*}{\textbf{Convex}}}
& \multirow{2}{*}{\begin{tabular}[c]{@{}c@{}}\textbf{Non-}\\ \textbf{convex}\end{tabular}} \\
& \multicolumn{1}{c|}{}& \multicolumn{1}{c|}{}& \multicolumn{1}{c|}{}&& \multicolumn{1}{c|}{}& \multicolumn{1}{c|}{}& \multicolumn{1}{c|}{}
&& \multicolumn{1}{c|}{}& \multicolumn{1}{c|}{}& \multicolumn{1}{c|}{}&\\
\hline
\multirow{1}{*}{TV \cite{li2009user}}
& \multicolumn{1}{c|}{\checkmark} & \multicolumn{1}{c|}{}&\multicolumn{1}{c|}{\checkmark}&&\multicolumn{1}{c|}{\checkmark}&\multicolumn{1}{c|}{}
& \multicolumn{1}{c|}{}&& \multicolumn{1}{c|}{\checkmark}& \multicolumn{1}{c|}{}& \multicolumn{1}{c|}{\checkmark}& \multicolumn{1}{c|}{}\\
\hline
\multirow{1}{*}{DL-CS \cite{Ravishankar2011MR}}
& \multicolumn{1}{c|}{\checkmark} & \multicolumn{1}{c|}{}&\multicolumn{1}{c|}{\checkmark}&&\multicolumn{1}{c|}{}&\multicolumn{1}{c|}{\checkmark}
& \multicolumn{1}{c|}{}&& \multicolumn{1}{c|}{\checkmark}& \multicolumn{1}{c|}{}& \multicolumn{1}{c|}{\checkmark}& \multicolumn{1}{c|}{}\\
\hline
\multirow{1}{*}{CSR \cite{zha2018compressed}}
&\multicolumn{1}{c|}{\checkmark}&\multicolumn{1}{c|}{}&\multicolumn{1}{c|}{\checkmark}&\multicolumn{1}{c|}{\checkmark}&\multicolumn{1}{c|}{}
&\multicolumn{1}{c|}{\checkmark}& \multicolumn{1}{c|}{}&& \multicolumn{1}{c|}{\checkmark}& \multicolumn{1}{c|}{}& \multicolumn{1}{c|}{\checkmark}& \multicolumn{1}{c|}{}\\
\hline
\multirow{1}{*}{CS-GSR \cite{Zhang2014Group}}
&\multicolumn{1}{c|}{\checkmark}&\multicolumn{1}{c|}{}&\multicolumn{1}{c|}{}&\multicolumn{1}{c|}{\checkmark}&\multicolumn{1}{c|}{}
&\multicolumn{1}{c|}{\checkmark}& \multicolumn{1}{c|}{\checkmark}& \multicolumn{1}{c|}{\checkmark}& \multicolumn{1}{c|}{\checkmark}& \multicolumn{1}{c|}{}& \multicolumn{1}{c|}{}& \multicolumn{1}{c|}{\checkmark}\\
\hline
\multirow{1}{*}{JASR \cite{Eslahi2016Compressive}}
&\multicolumn{1}{c|}{\checkmark}&\multicolumn{1}{c|}{}&\multicolumn{1}{c|}{}&\multicolumn{1}{c|}{\checkmark}&\multicolumn{1}{c|}{\checkmark}
&\multicolumn{1}{c|}{}& \multicolumn{1}{c|}{\checkmark}& \multicolumn{1}{c|}{}& \multicolumn{1}{c|}{\checkmark}& \multicolumn{1}{c|}{}& \multicolumn{1}{c|}{\checkmark}& \multicolumn{1}{c|}{}\\
\hline
\multirow{1}{*}{GSRC \cite{Zha2020Group}}
&\multicolumn{1}{c|}{\checkmark}&\multicolumn{1}{c|}{}&\multicolumn{1}{c|}{}&\multicolumn{1}{c|}{\checkmark}&\multicolumn{1}{c|}{}
&\multicolumn{1}{c|}{\checkmark}& \multicolumn{1}{c|}{\checkmark}& \multicolumn{1}{c|}{}& \multicolumn{1}{c|}{\checkmark}& \multicolumn{1}{c|}{}& \multicolumn{1}{c|}{\checkmark}& \multicolumn{1}{c|}{}\\
\hline
\multirow{1}{*}{GSC-NCR \cite{Zha2020Benchmark}}
&\multicolumn{1}{c|}{\checkmark}&\multicolumn{1}{c|}{}&\multicolumn{1}{c|}{}&\multicolumn{1}{c|}{\checkmark}&\multicolumn{1}{c|}{}
&\multicolumn{1}{c|}{\checkmark}& \multicolumn{1}{c|}{\checkmark}& \multicolumn{1}{c|}{}& \multicolumn{1}{c|}{\checkmark}& \multicolumn{1}{c|}{}& \multicolumn{1}{c|}{}& \multicolumn{1}{c|}{\checkmark}\\
\hline
\multirow{1}{*}{HSSE \cite{Zha2021Hybrid}}
&\multicolumn{1}{c|}{\checkmark}&\multicolumn{1}{c|}{\checkmark}&\multicolumn{1}{c|}{}&\multicolumn{1}{c|}{\checkmark}&\multicolumn{1}{c|}{}
&\multicolumn{1}{c|}{\checkmark}& \multicolumn{1}{c|}{\checkmark}& \multicolumn{1}{c|}{}& \multicolumn{1}{c|}{\checkmark}& \multicolumn{1}{c|}{}& \multicolumn{1}{c|}{\checkmark}& \multicolumn{1}{c|}{}\\
\hline
\multirow{1}{*}{CS-NLR \cite{Dong2014Compressive}}
&\multicolumn{1}{c|}{\checkmark}&\multicolumn{1}{c|}{}&\multicolumn{1}{c|}{}&\multicolumn{1}{c|}{\checkmark}&\multicolumn{1}{c|}{}
&\multicolumn{1}{c|}{\checkmark}& \multicolumn{1}{c|}{}& \multicolumn{1}{c|}{\checkmark}& \multicolumn{1}{c|}{\checkmark}& \multicolumn{1}{c|}{}& \multicolumn{1}{c|}{}& \multicolumn{1}{c|}{\checkmark}\\
\hline
\multirow{1}{*}{RRC \cite{Zha2020From}}
&\multicolumn{1}{c|}{\checkmark}&\multicolumn{1}{c|}{}&\multicolumn{1}{c|}{}&\multicolumn{1}{c|}{\checkmark}&\multicolumn{1}{c|}{}
&\multicolumn{1}{c|}{\checkmark}& \multicolumn{1}{c|}{}& \multicolumn{1}{c|}{\checkmark}& \multicolumn{1}{c|}{\checkmark}& \multicolumn{1}{c|}{}& \multicolumn{1}{c|}{\checkmark}& \multicolumn{1}{c|}{}\\
\hline
\multirow{1}{*}{LR-GSC \cite{Zha2021Reconciliation}}
&\multicolumn{1}{c|}{\checkmark}&\multicolumn{1}{c|}{}&\multicolumn{1}{c|}{}&\multicolumn{1}{c|}{\checkmark}&\multicolumn{1}{c|}{}
&\multicolumn{1}{c|}{\checkmark}& \multicolumn{1}{c|}{\checkmark}& \multicolumn{1}{c|}{\checkmark}& \multicolumn{1}{c|}{\checkmark}& \multicolumn{1}{c|}{}& \multicolumn{1}{c|}{\checkmark}& \multicolumn{1}{c|}{}\\
\hline
\multirow{1}{*}{CS-Net \cite{Shi2020Image}}
&\multicolumn{1}{c|}{}&\multicolumn{1}{c|}{\checkmark}&\multicolumn{1}{c|}{\checkmark}&\multicolumn{1}{c|}{}&\multicolumn{1}{c|}{}
&\multicolumn{1}{c|}{\checkmark}& \multicolumn{1}{c|}{}& \multicolumn{1}{c|}{}& \multicolumn{1}{c|}{}& \multicolumn{1}{c|}{\checkmark}& \multicolumn{1}{c|}{}& \multicolumn{1}{c|}{\checkmark}\\
\hline
\multirow{1}{*}{NL-CSNet \cite{Cui2021Image}}
&\multicolumn{1}{c|}{}&\multicolumn{1}{c|}{\checkmark}&\multicolumn{1}{c|}{}&\multicolumn{1}{c|}{\checkmark}&\multicolumn{1}{c|}{}
&\multicolumn{1}{c|}{\checkmark}& \multicolumn{1}{c|}{}& \multicolumn{1}{c|}{}& \multicolumn{1}{c|}{}& \multicolumn{1}{c|}{\checkmark}& \multicolumn{1}{c|}{}& \multicolumn{1}{c|}{\checkmark}\\
\hline

\end{tabular}
}
\label{Tab:1}
\vspace{-8mm}
\end{table*}

\vspace{-0.15in}
\subsection{Qualitative Comparison of Different Image CS Methods}
\label{sec:2.3}

Table~\ref{Tab:1} presents a qualitative comparison of a sample set of methods based on priors, models, and properties they utilize. It can be seen that TV \cite{li2009user}, DL-CS \cite{Ravishankar2011MR}, and CS-Net \cite{Shi2020Image} methods exploit local image priors, while other methods consider image NSS priors. In particular, the HSSE method  \cite{Zha2021Hybrid} uses  both internal and external (\ie, natural image dataset) NSS priors. From the perspective of models, TV \cite{li2009user} and JASR \cite{Eslahi2016Compressive} methods are based on analytical models, while other methods are learning-based models. CS-GSR \cite{Zhang2014Group} and LR-GSC \cite{Zha2021Reconciliation} methods both integrate GSR and LR models. From the perspective of model properties, deep learning methods (\ie, CS-Net \cite{Shi2020Image} and NL-CSNet \cite{Cui2021Image}) are supervised learning schemes, while other methods are self-supervised learning schemes.

\vspace{-0.15in}
\section{Tutorial on Nonlocal Sparse and LR Models for Image CS}
\label{sec:3}

Nonlocal structured sparsity prior using GSR and/or LR modeling has been shown to be effective in reconstructing images from limited measurements. Since various GSR- and LR-based algorithms have been proposed for image CS, each based on different models (\ie, GSR or LR or both of them) and learning schemes, it is crucial to answer the following questions:

\begin{enumerate}
\item How to construct a general variational formulation of GSR and LR learning for image CS?
\item What are the image properties been exploited in each nonlocal sparse or LR learning model?
\item What are the relationships and differences between the GSR and LR schemes?
\item Which method is the most effective for reconstructing an image?
\end{enumerate}

Toward this end, we present a tutorial that aims to unify all recent GSR and LR schemes and summarize their problem formulations and algorithms based on using a general framework. We further discuss and compare the characteristics of several GSR and LR approaches, including CS-GSR \cite{Zhang2014Group}, GSRC \cite{Zha2020Group}, HSSE \cite{Zha2021Hybrid}, CS-NLR \cite{Dong2014Compressive}, RRC \cite{Zha2020From}, and LR-GSC \cite{Zha2021Reconciliation}. Moreover, we investigate the relationship between the GSR and LR models and illustrate the effectiveness of GSR- and LR-based image CS methods, \eg, using HSSE \cite{Zha2021Hybrid} compared to other classes of image CS reconstruction methods.

\vspace{-0.15in}
\subsection{Image CS Formulation}
\label{sec:3.1}

CS theory \cite{Candes2008An} states that when a signal is sparse in some domain, it can be accurately reconstructed from much fewer linear measurements than required by the Nyquist-Shannon sampling theorem. For image CS, one strives for the accurate reconstruction of a vectorized image $\textbf{\emph{x}} \in{\mathbb R}^{N}$ from its limited measurements $\textbf{\emph{y}} \in{\mathbb R}^{M}$, namely, $\textbf{\emph{y}} = \boldsymbol\Phi\textbf{\emph{x}}$, where $\boldsymbol\Phi\in{\mathbb R}^{M\times N}$ represents a measurement matrix, where $M$ is much smaller than $N$. In order to recover the image $\textbf{\emph{x}}$ from its limited measurements $\textbf{\emph{y}}$, prior knowledge about $\textbf{\emph{x}}$ is required. The standard CS method of reconstructing an image $\textbf{\emph{x}}$ from its limited measurements $\textbf{\emph{y}}$ is formulated as the following constrained minimization problem:
{\setlength\abovedisplayskip{4pt}
\setlength\belowdisplayskip{4pt}
\begin{equation}
{(\rm P0)    \ \ \ \ \ \ \ \hat{\textbf{\emph{x}}} =\argmin_{\textbf{\emph{x}}} \|{{\bf\Psi}}\textbf{\emph{x}}\|_0,  \ \ \ { s.t.}   \ \ \ \textbf{\emph{y}} = {\bf\Phi}\textbf{\emph{x}}, \nonumber}
\label{eq:1}
\end{equation}}%
where $\boldsymbol\Psi$ represents a sparsifying transform space, and $\textbf{\emph{x}}$ is assumed to be sparse in the $\boldsymbol\Psi$ transform domain. The operation $\|~\|_0$ denotes the $\ell_0$ ``norm" sparsity measure, which counts the number of nonzero entries in a vector. Alternative sparsity-promoting penalty functions include the convex $\ell_1$ norm or the non-convex $\ell_p$ norm (0$<p<$1). (P0) aims to reconstruct the unknown image by seeking the solution that is the sparsest in the $\boldsymbol\Psi$ domain while satisfying $\textbf{\emph{y}} = \boldsymbol\Phi\textbf{\emph{x}}$. However, since the sampled measurements $\textbf{\emph{y}}$ are typically noisy, the image CS reconstruction problem is usually formulated in a maximum a posteriori (MAP) framework \cite{kotera2013blind} with a data fidelity penalty as
{\setlength\abovedisplayskip{4pt}
\setlength\belowdisplayskip{4pt}
\begin{equation}
{(\rm P1)  \ \ \ \ \ \ \ \hat{\textbf{\emph{x}}}=\argmin\limits_{\textbf{\emph{x}}}
\frac{1}{2}\left\|{\textbf{\emph{y}}}-{\bf\Phi}\textbf{\emph{x}}\right\|_2^2+ \lambda\|{\bf\Psi}\textbf{\emph{x}}\|_0,}
\nonumber
\label{eq:2}
\end{equation}}%
where $\frac{1}{2}\left\|{\textbf{\emph{y}}}-\boldsymbol\Phi\textbf{\emph{x}}\right\|_2^2$ denotes the data fidelity term based on a Gaussian measurement noise model, $\|~\|_2$ denotes the $\ell_2$ norm, and $\lambda >0$ is a regularization parameter.

\vspace{-0.15in}
\subsection{A Generalized Nonlocal Structured Sparsity Framework for Image CS and Its Variations}
\label{sec:3.2}
In practice, there are several limitations for applying (P0) or (P1) for image reconstruction:
\begin{itemize}
\item Imposing a common sparsity model for the entire image is restrictive for modeling diverse textures  \cite{Zhang2014Image}.
\item The transform $\boldsymbol\Psi$ is predefined and fixed, thus may not be optimal for the underlying image $\hat{\textbf{\emph{x}}}$ \cite{Zha2021Hybrid}.
\item They both involve difficult combinatorial optimization problems, which are NP-hard in general \cite{Dong2014Compressive}.
\end{itemize}

\begin{figure*}[!t]
\vspace{-2mm}
\centering
\begin{minipage}[b]{1\linewidth}
{\includegraphics[width = 1\textwidth]{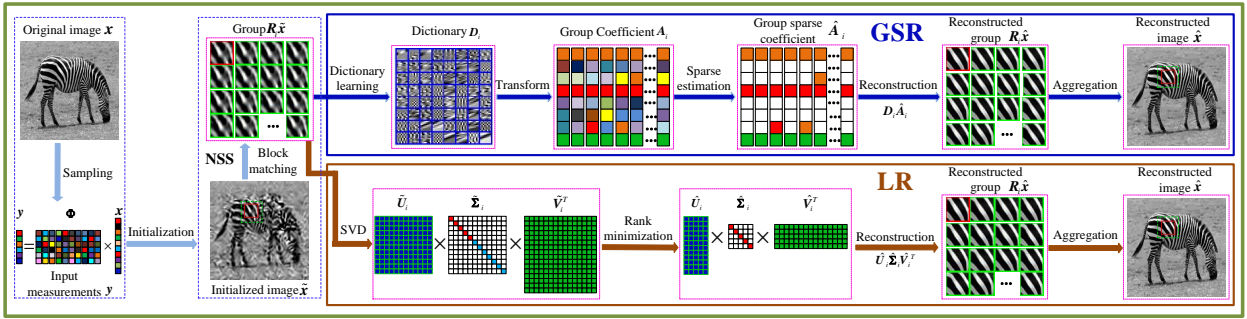}}
\end{minipage}
\vspace{-6mm}
\caption{The diagram of the general image CS pipeline using the nonlocal structured sparsity priors by GSR and LR modeling.}
\label{fig:4}
\vspace{-8mm}
\end{figure*}

Recent advances in nonlocal structured sparsity approaches have shown great potential in addressing these limitations by exploiting strong correlations among nonlocal similar patches and modeling these similar patches using structured sparsity priors. Such schemes can significantly reduce the uncertainty in reconstruction of unknown signals and produce more accurate results by solving
{\setlength\abovedisplayskip{4pt}
\setlength\belowdisplayskip{4pt}
\begin{equation}
{(\rm P2)  \ \ \ \ \ \ \ \hat{\textbf{\emph{x}}}=\argmin\limits_{\textbf{\emph{x}}}
\frac{1}{2}\left\|{\textbf{\emph{y}}}-\bf\Phi\textbf{\emph{x}}\right\|_2^2+ \lambda \boldsymbol\Re_{NL} (\textbf{\emph{x}}),}
\nonumber
\label{eq:3}
\end{equation}}%
where $\boldsymbol\Re_{\rm NL} (\textbf{\emph{x}})$ denotes a nonlocal structured sparsity regularizer \cite{Zhang2014Group}. The exact form of $\boldsymbol\Re_{\rm NL} (\textbf{\emph{x}})$ depends on the actual nonlocal image model being applied, which constructs the main difference amongst the various types of nonlocal structured sparsity formulations. One can integrate the desired nonlocal structured sparsity prior through $\boldsymbol\Re_{\rm NL} (\textbf{\emph{x}})$ into (P2), leading to different nonlocal image CS schemes \cite{Dong2014Compressive,Zha2020Group,Zha2021Hybrid}. The corresponding algorithms for solving the variation of (P2) can be summarized in the following three steps:
\begin{enumerate} [1)]
    \item Compute the similarity between each exemplar patch and the candidate patches of a nonlocal region using a variant of the $K$-nearest-neighbor (KNN) algorithm. The most similar patches are grouped by the block matching operator~\cite{Dabov2007Image,Gu2014Weighted}, ranked by their Euclidean distance to the exemplar patch.
    \item Each group of selected patches forms the data matrix (see Fig.~\ref{fig:1}), by stacking the \emph{vectorized} patches. Apply the corresponding nonlocal modeling to reconstruct each data matrix.
    \item The reconstructed patches are reshaped and deposit back to the corresponding locations of the image, followed by the image update process to recover $\hat{\textbf{\emph{x}}}$.
\end{enumerate}

Fig.~\ref{fig:4} shows a general pipeline for image CS using the nonlocal structural sparsity prior via GSR or LR modeling. We next review some recent GSR and LR approaches that exploit the nonlocal structured sparsity prior for image CS, including CS-GSR \cite{Zhang2014Group}, GSRC \cite{Zha2020Group}, HSSE \cite{Zha2021Hybrid}, CS-NLR \cite{Dong2014Compressive}, RRC \cite{Zha2020From}, and LR-GSC \cite{Zha2021Reconciliation}. We discuss how to incorporate them into the general framework (P2) by leveraging specific models, properties, and learned regularizers. We summarize the methods under the more common blind image CS setup \cite{Ravishankar2011MR,Zhang2014Image}, which simultaneously learns the model with reconstruction.
\vspace{-0.15in}
\subsection{GSR-based Methods for Image CS}
\label{sec:3.3}

\begin{figure*}[!t]
  \vspace{-2mm}
  \centering
  \begin{minipage}[b]{0.65\linewidth}
{\includegraphics[width= 1\textwidth]{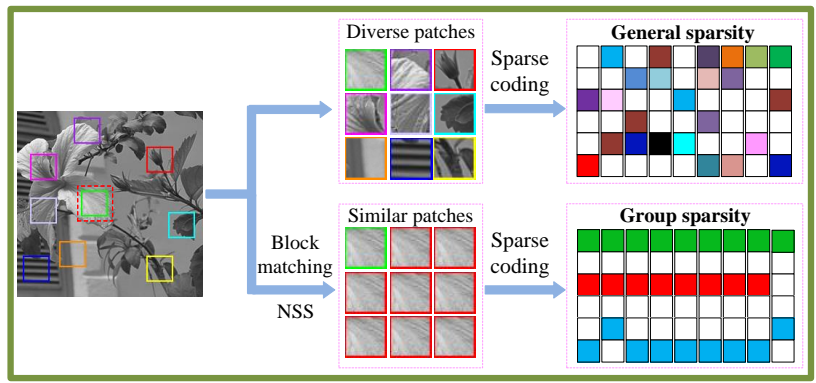}}
\end{minipage}
\vspace{-4mm}
\caption{A comparison between general sparsity (the columns in the representation are sparse but with quite different patterns) and group sparsity (each column is sparse and sufficiently similar).}
\label{fig:5}
\vspace{-8mm}
\end{figure*}

Different from the conventional
sparse model that uses each patch as the basic unit for sparse representation (or sparse coding), the GSR model employs each group containing many nonlocal similar patches as the basic unit for sparse representation. It employs an image NSS prior to capture the strong correlations of nonlocal similar patches. Fig.~\ref{fig:5} displays the difference between general sparsity and group sparsity (color boxes represent non-zero elements and white boxes represent zeros).

\vspace{-0.05in}
\subsubsection{CS-GSR}
\label{sec:3.3.1}

The formulation of the classic GSR model for image CS is dubbed CS-GSR \cite{Zhang2014Group}. The regularizer $\boldsymbol\Re_{\rm NL} (\textbf{\emph{x}}) = \boldsymbol\Re_{\rm CS-GSR} (\textbf{\emph{x}})$ is defined as
{\setlength\abovedisplayskip{4pt}
\setlength\belowdisplayskip{6pt}
\begin{equation}
{\boldsymbol\Re_{\rm CS-GSR} (\textbf{\emph{x}}) \triangleq  \argmin\limits_{\{\textbf{\emph{A}}_i\}} \frac{1}{2\mu}\sum\nolimits_{i=1}^n \{ \|{\textbf{\emph{R}}}_i\textbf{\emph{x}}-{\textbf{\emph{D}}}_i{\textbf{\emph{A}}}_i\|_F^2+\lambda\|\textbf{\emph{A}}_i\|_0\}.}
\label{eq:1.1}
\end{equation}
Here and in the rest of this article, when some indexed variable is enclosed within curly brackets, it represents the set of all variables over all index values, \eg, $\{{\textbf{\emph{A}}}_i\}$ represents $\{{\textbf{\emph{A}}}_i\}_{i=1}^n$. ${\textbf{\emph{R}}}_i$ denotes a KNN operator that selects the $m$ closest patches to the $i$-th exemplar patch $\textbf{\emph{x}}_i\in{\mathbb R}^{\sqrt{b}\times \sqrt{b}}$ in the image $\textbf{\emph{x}}$, and forms a group ${\textbf{\emph{X}}}_i\in{\mathbb R}^{b\times m}$, \ie, ${\textbf{\emph{R}}}_i\textbf{\emph{x}} = {\textbf{\emph{X}}}_i = [\textbf{\emph{x}}_{i,1}, \textbf{\emph{x}}_{i,2},\dots, \textbf{\emph{x}}_{i,m}]$, and $\textbf{\emph{x}}_{i,j}$ denotes the $j$-th patch in the $i$-th group ${\textbf{\emph{X}}}_i$. In general GSR models, the dictionary ${\textbf{\emph{D}}}_i$ is usually learned from each group ${\textbf{\emph{X}}}_i$, such as a principal component analysis (PCA) dictionary \cite{Zha2021Hybrid} or graph dictionary \cite{Cheung2018Graph}. Note that instead of using a fixed analytical dictionary, the dictionary is updated from the reconstructed group $\hat{\textbf{\emph{X}}}_i = {\textbf{\emph{D}}}_i\hat{\textbf{\emph{A}}}_i$ in each iteration using PCA \cite{Zha2021Hybrid}, singular value decomposition (SVD) \cite{Zhang2014Group}, graph decomposition \cite{Cheung2018Graph}, etc. $\hat{\textbf{\emph{A}}}_i$ represents the group sparse coefficient for each group ${\textbf{\emph{X}}}_i$. Also, $\|\cdot \|_F$ denotes the Frobenius norm and the $\ell_0$ ``norm" is used to measure the true sparsity of ${\textbf{\emph{A}}}_i$ to improve the quality of the reconstructed image. $\mu$ is a balancing factor that makes the formulation (P2) more tractable.

In the CS-GSR model \cite{Zhang2014Group}, a self-adaptive dictionary is learned from each group ${\textbf{\emph{X}}}_i$ by applying LR SVD for the group ${\textbf{\emph{X}}}_i$, therefore leading to the model~\eqref{eq:1.1} with both group sparsity and LR properties. Moreover, although solving the formulation of the CS-GSR model in~\eqref{eq:1.1} is a complex combinatorial optimization problem,  a convex optimization algorithm framework is employed to solve~\eqref{eq:1.1}, which obtains an effective solution and good convergence behavior.

\vspace{-0.05in}
\subsubsection{GSRC}
\label{sec:3.3.3}
Most of the existing GSR-based models measure the sparsity of $\textbf{\emph{A}}_i$ by directly conducting the norm minimization for $\textbf{\emph{A}}_i$, such as  $\|\textbf{\emph{A}}_i\|_1$ and $\|\textbf{\emph{A}}_i\|_0$, which have been demonstrated effectiveness in image CS \cite{Zhang2014Group,Zha2020Benchmark}. However, when the sampling rate is low, such methods may not faithfully estimate the sparsity of each group and thus lead to degraded image reconstruction quality. Recent work proposed to use GSR with an auxiliary residual prior for image CS. This is termed GSRC \cite{Zha2020Group} and is formulated with the following regularizer:
{\setlength\abovedisplayskip{4pt}
\setlength\belowdisplayskip{6pt}
\begin{equation}
{\boldsymbol\Re_{\rm GSRC} (\textbf{\emph{x}}) \triangleq  \argmin\limits_{\{\textbf{\emph{A}}_i\}} \frac{1}{2\mu}\sum\nolimits_{i=1}^n \{ \|{\textbf{\emph{R}}}_i\textbf{\emph{x}}-{\textbf{\emph{D}}}_i{\textbf{\emph{A}}}_i\|_F^2+\lambda\|\textbf{\emph{A}}_i - \textbf{\emph{B}}_i\|_1\},}
\label{eq:3}
\end{equation}}
where ${\textbf{\emph{B}}}_i$ represents the group coefficient of the original group, which is derived from the original image ${\textbf{\emph{x}}}$. The principle of the GSRC model is that each degraded group sparse coefficient $\hat{\textbf{\emph{A}}}_i$ is expected to be close enough to the corresponding original group sparse coefficient $\hat{\textbf{\emph{B}}}_i$. Therefore, image reconstruction quality largely depends on the group sparsity residual $\textbf{\emph{P}}_i = \textbf{\emph{A}}_i- \textbf{\emph{B}}_i$. In order to achieve good performance in image CS, the group sparsity residual $\textbf{\emph{P}}_i$ for each group should be suppressed as far as possible. However, the real group coefficient ${\textbf{\emph{B}}}_i$ does not exist as the original image is unavailable. To this end, the GSRC approach designs a self-supervised learning scheme to estimate ${\textbf{\emph{B}}}_i$ directly from input data $\textbf{\emph{X}}_i ({\textbf{\emph{R}}}_i\textbf{\emph{x}})$. Specifically, inspired by NLM for image denoising \cite{Buades2005A}, for each group $\textbf{\emph{X}}_i$ containing $m$ nonlocal similar patches, ${\textbf{\emph{X}}}_i = [\textbf{\emph{x}}_{i,1}, \textbf{\emph{x}}_{i,2},\dots, \textbf{\emph{x}}_{i,m}]$, a good estimate of $\boldsymbol \beta$ can be obtained by computing the weighted average of each vector $\boldsymbol \alpha_{j}$ in  ${\textbf{\emph{A}}}_i$, namely, $\boldsymbol \beta = \sum\nolimits_{j=1}^m \textbf{\emph{w}}_{j}\boldsymbol \alpha_{j}$, where the vector $\boldsymbol \alpha_{j}$ denotes the $j$-th  vector of   $\textbf{\emph{A}}_i$. ${\textbf{\emph{A}}}_i = {\textbf{\emph{D}}}_i^{T}\textbf{\emph{X}}_i$ and 
the PCA sub-dictionary ${\textbf{\emph{D}}}_i$ is estimated from each group $\textbf{\emph{X}}_i$ \cite{Zha2021Hybrid}. Scalar $\textbf{\emph{w}}_{j}$ represents a weight that is inversely proportional to the distance between the exemplar patch ${\textbf{\emph{x}}}_i$ (of $\textbf{\emph{X}}_i$) and its similar patch ${\textbf{\emph{x}}}_{i,j}$, \ie, $\textbf{\emph{w}}_{j}= \frac{1}{L}{\rm exp}(-\left\|\textbf{\emph{x}}_{i}-\textbf{\emph{x}}_{i,j}\right\|_2^2/h)$, where $h$ represents a predefined constant and $L$ represents a normalization factor \cite{Buades2005A}. Then, by simply replicating $\boldsymbol\beta$ $m$ times as the columns of $\textbf{\emph{B}}_i$, namely, $\textbf{\emph{B}}_i = [ \boldsymbol \beta_{1},  \boldsymbol \beta_{2}, ...,  \boldsymbol \beta_{m}]$, where each column of $\textbf{\emph{B}}_i$ is equal to  $\boldsymbol \beta$. The GSRC model manifests that integrating an auxiliary residual prior into the classic GSR model can enable estimating the unknown group sparse coefficients  $\{\hat{\textbf{\emph{A}}}_i\}_{i=1}^n$ more accurately, and thus achieve better reconstruction performance than classic GSR-based methods.

\vspace{-0.05in}
\subsubsection{HSSE}
\label{sec:3.3.4}

The aforementioned variations of GSR-based image CS schemes \cite{Zhang2014Group,Zha2020Group} only model the sparsity of the image by exploiting internal NSS prior of the image itself, while ignoring the beneficial external NSS prior from natural images, which is complementary to the internal NSS prior. The recently proposed HSSE model \cite{Zha2021Hybrid} jointly employs image NSS prior using both the internal and external image data that provide complementary information for reconstruction. The corresponding HSSE employs the regularizer $\boldsymbol\Re_{\rm NL} (\textbf{\emph{x}}) = \boldsymbol\Re_{\rm HSSE} (\textbf{\emph{x}})$, which contains two complementary parts, namely,
{\setlength\abovedisplayskip{4pt}
\setlength\belowdisplayskip{6pt}
\begin{equation}
{\boldsymbol\Re_{\rm HSSE} (\textbf{\emph{x}}) \triangleq  \argmin\limits_{\{\textbf{\emph{A}}_i\}, \{\textbf{\emph{B}}_i\}}\sum\nolimits_{i=1}^n \{  \frac{1}{2\mu}\|{\textbf{\emph{R}}}_i \textbf{\emph{x}} -{\textbf{\emph{D}}}_i{\textbf{\emph{A}}}_i\|_F^2 + \lambda\|{\textbf{\emph{A}}}_i\|_1  +\frac{1}{2\rho}\|{\textbf{\emph{U}}}_i{\textbf{\emph{B}}}_i-{\textbf{\emph{D}}}_i{\textbf{\emph{A}}}_i\|_F^2+ \tau\|{\textbf{\emph{B}}}_i\|_1\},}
\label{eq:6}
\end{equation}}
where the first two terms exploit an internal NSS prior, while the latter two terms employ an external NSS prior. Here, ${\textbf{\emph{U}}}_i$ is an external sub-dictionary that is learned from the image groups
(using external NSS prior)
of the external image corpus, and ${\textbf{\emph{B}}}_i$ represents the external group coefficient. $\tau$ is a regularization parameter, and $\rho$ plays a role similar to parameter $\mu$ of~\eqref{eq:1.1}. The internal sub-dictionary ${\textbf{\emph{D}}}_i$ is learned from the corresponding group via the PCA operation in each iteration. For the external sub-dictionary ${\textbf{\emph{U}}}_i$, many groups are constructed via the blocking matching operator \cite{Dabov2007Image,Gu2014Weighted} from the external image dataset, and then a group-based  (GMM) learning algorithm \cite{Xu2015Patch} is developed to learn many PCA sub-dictionaries $\{{\textbf{\emph{U}}}_i\}_{i=1}^M$ from these constructed external groups, where $M$ represents the total number of Gaussian components. Finally, the best matched PCA sub-dictionary ${\textbf{\emph{U}}}_i$ is selected for group  $\textbf{\emph{R}}_i\textbf{\emph{x}}$. The goal of HSSE is to represent an image by jointly employing the NSS prior of both the internal image and external image corpus, which can effectively alleviate the overfitting to image data degradation by incorporating (P2) and thus produces superior reconstruction results.

\vspace{-0.15in}
\subsection{Nonlocal LR-based Methods for Image CS}
\label{sec:3.4}
Alternatively, through using the NSS prior, many similar patches are clustered into a group matrix containing $m$ similar patches, \ie, ${\textbf{\emph{R}}}_i\textbf{\emph{x}} = {\textbf{\emph{X}}}_i = [\textbf{\emph{x}}_{i,1}, \textbf{\emph{x}}_{i,2},\dots, \textbf{\emph{x}}_{i,m}]\in{\mathbb R}^{b\times m}$, and $\textbf{\emph{x}}_{i,m}\in{\mathbb R}^{b}$ denotes the $m$-th patch in the $i$-th group ${\textbf{\emph{X}}}_i$, This group matrix ${\textbf{\emph{R}}}_i\textbf{\emph{x}}$ has LR property because each of its columns has a similar structure. Therefore, the nonlocal structured sparsity prior using LR modeling has also shown great potential in image CS. The LR approach generally involves applying an SVD to the group matrix ${\textbf{\emph{R}}}_i\textbf{\emph{x}}$ and using a rank minimization algorithm to maintain the LR components while discarding the non-LR components in ${\textbf{\emph{R}}}_i \textbf{\emph{x}}$, as shown in Fig.~\ref{fig:6}.
The most representative rank minimization algorithm is nuclear norm minimization (NNM) \cite{cai2010singular}, which aims to find a LR matrix $\textbf{\emph{Z}}_i$ with rank $r \ll {\rm min (b, m)}$ from ${\textbf{\emph{X}}}_i$ by solving
{\setlength\abovedisplayskip{4pt}
\setlength\belowdisplayskip{1pt}
\begin{equation}
{\hat{\textbf{\emph{Z}}}_i = \arg\min_{\textbf{\emph{Z}}_i}  \frac{1}{2}\{\left\|{\textbf{\emph{X}}}_i -\textbf{\emph{Z}}_i\right\|_F^2 +\lambda \left\|\textbf{\emph{Z}}_i\right\|_*\} \; \; \; \forall i \;,}
\label{eq:7}
\end{equation}}
where $\|\cdot\|_*$  represents the nuclear norm  \cite{cai2010singular} ($\left\|\textbf{\emph{Z}}_i\right\|_* = \sum_j\boldsymbol\sigma_{i,j}$, with $\boldsymbol\sigma_{i,j}$ denoting the $j$-th singular value of the matrix $\textbf{\emph{Z}}_i$) and $\lambda$ is a positive constant. While singular value thresholding (SVT)  \cite{cai2010singular} provides good theoretical guarantees for NNM, all singular values are shrunk equally by the NNM algorithm, which neglects the different significance of different singular values and thus leads to an unsatisfactory matrix rank estimation. Meanwhile, the NNM algorithm is prone to over-shrink the matrix rank components (\ie, deviated from the original rank components), thus limiting its capability in various image reconstruction applications. To overcome the limitations of the NNM algorithm, most recently, many excellent rank minimization algorithms, such as WNNM \cite{Gu2014Weighted} and rank residual constraint (RRC) \cite{Zha2020From} have been developed for computational imaging settings.

\begin{figure*}[!t]
  \vspace{-2mm}
  \centering
  \begin{minipage}[b]{0.65\linewidth}
{\includegraphics[width= 1\textwidth]{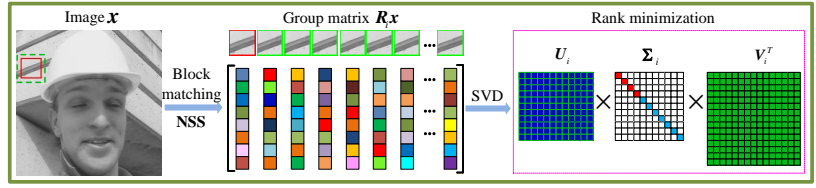}}
\end{minipage}
\vspace{-2mm}
\caption{By applying SVD to the group matrix ${\textbf{\emph{R}}}_i \textbf{\emph{x}}$, LR approach employes the rank minimization algorithm to preserve the LR components (red boxes in $\boldsymbol\Sigma_i$), while discarding non-LR components (light blue boxes in $\boldsymbol\Sigma_i$).}
\label{fig:6}
\vspace{-8mm}
\end{figure*}

\vspace{-0.05in}
\subsubsection{CS-NLR}
\label{sec:3.4.1}
It is well-known that when the singular values of the matrix are arranged in descending order, then the larger singular values are often used to quantify information of its underlying principal directions. Motivated by this, WNNM \cite{Gu2014Weighted} assigns different weights to different singular values to estimate the matrix rank more accurately.
The CS-NLR approach \cite{Dong2014Compressive} applies WNNM as the regularizer, with $\boldsymbol\Re_{\rm NL} (\textbf{\emph{x}}) = \boldsymbol\Re_{\rm CS-NLR} (\textbf{\emph{x}})$ as follows:
{\setlength\abovedisplayskip{4pt}
\setlength\belowdisplayskip{6pt}
\begin{equation}
{\boldsymbol\Re_{\rm CS-NLR} (\textbf{\emph{x}}) \triangleq \arg\min_{\{\textbf{\emph{Z}}_i\}}  \frac{1}{2\mu}\sum\nolimits_{i=1}^n \{\left\|{\textbf{\emph{R}}}_i \textbf{\emph{x}} -\textbf{\emph{Z}}_i\right\|_F^2 +\lambda \left\|\textbf{\emph{Z}}_i\right\|_{\textbf{\emph{w}}_i, *}\} ,}
\label{eq:8}
\end{equation}}
where $\left\|\textbf{\emph{Z}}_i\right\|_{\textbf{\emph{w}}_i, *} = \sum_j w_{i,j}\boldsymbol\sigma_{i,j}$, and $\boldsymbol\sigma_{i,j}$ is the $j$-th singular value of $\textbf{\emph{Z}}_i$. $\textbf{\emph{w}}_i = \{w_{i,1}, \dots, w_{i, g}\}, g = {\rm min} (b,m)$. Here, $w_{i, j}\geq0$ is the weight assigned to the corresponding $\boldsymbol\sigma_{i,j}$, which is usually set to $w_{i, j} = k/(\boldsymbol\sigma_{i,j} + \epsilon)$, where $k$ and $\epsilon$ are a non-negative constant and a small constant to avoid dividing by zero, respectively. It is worth noting that the weights are ranked in ascending order since the singular values are arranged in descending order. In addition,  the initial value of $\boldsymbol\sigma_{i,j}$ is set to the singular values of ${\textbf{\emph{R}}}_i \textbf{\emph{x}}$ in the iteration algorithm. Overall, compared with NNM \cite{cai2010singular}, the CS-NLR method applies WNNM as the regularizer, which explicitly considers physical meanings of the singular values (\ie, singular values are ranked in descending order),  making the matrix rank estimation more accurate and thus leading to better CS reconstruction results.

\vspace{-0.05in}
\subsubsection{RRC}
\label{sec:3.4.2}

The classic NNM tends to over-shrink the rank components, namely, the singular values of the recovered matrix deviate from the singular values of the original matrix. However, in LR matrix estimation, the singular values of the recovered matrix are expected to be as close as possible to the singular values of the original matrix. To this end, the RRC approach \cite{Zha2020From} progressively approximates the underlying LR matrix via minimizing the rank residual, which applies the regularizer $\boldsymbol\Re_{\rm NL} (\textbf{\emph{x}}) = \boldsymbol\Re_{\rm RRC} (\textbf{\emph{x}})$ defined as,
{\setlength\abovedisplayskip{4pt}
\setlength\belowdisplayskip{6pt}
\begin{equation}
{\boldsymbol\Re_{\rm RRC} (\textbf{\emph{x}}) \triangleq \arg\min_{\{\textbf{\emph{Z}}_i\}}  \frac{1}{2\mu}\sum\nolimits_{i=1}^n \{\left\|{\textbf{\emph{R}}}_i \textbf{\emph{x}} -\textbf{\emph{Z}}_i\right\|_F^2 +\lambda \left\|\boldsymbol\gamma_i\right\|_1\} ,}
\label{eq:9}
\end{equation}}
where $\boldsymbol\gamma_i \stackrel{\rm def}{=}\boldsymbol\sigma_i-\boldsymbol\psi_i$ with $\boldsymbol\sigma_i$ being the singular value of $\textbf{\emph{Z}}_i$. $\boldsymbol\psi_i$ should ideally be the singular value of the $i$-th original matrix, but the original matrix is not available in practice. Therefore, $\boldsymbol\psi_i$ is defined as the singular value of the $i$-th reference matrix $\textbf{\emph{X}}_i'$, where $\textbf{\emph{X}}_i'$ is a good estimate of the original matrix. In image CS, the estimated reference matrix $\textbf{\emph{X}}_i'$ depends on the prior knowledge of the original image. Inspired by the fact that natural images often possess many repetitive structures \cite{Dabov2007Image}, the RRC approach employs image NSS prior using nonlocal similar patches to estimate the reference matrix $\textbf{\emph{X}}_i'$. To be concrete, a good estimate of each reference patch ${\textbf{\emph{x}}}_{i,j}'$  in $\textbf{\emph{X}}_i'$ can be computed by the weighted average of  the patches $\{\hat{\textbf{\emph{x}}}_{i, k}\}$ in ${\textbf{\emph{R}}}_i \textbf{\emph{x}}$, for each group ${\textbf{\emph{R}}}_i \textbf{\emph{x}}$ containing $m$ nonlocal similar patches, \ie, ${\textbf{\emph{x}}}_{i,j}' =\sum\nolimits_{k=1}^{m-j+1} \textbf{\emph{w}}_{i,k}{\textbf{\emph{x}}}_{i,k}$, which is based on NLM filtering \cite{Buades2005A}. Here, ${\textbf{\emph{x}}}_{i,j}'$ and $\hat{\textbf{\emph{x}}}_{i,k}$ denote the $j$-th and $k$-th patch of $\textbf{\emph{X}}_i'$ and ${\textbf{\emph{R}}}_i \textbf{\emph{x}}$, respectively. $\textbf{\emph{w}}_{i,k}$ denotes the weight, which is inversely proportional to the distance between the exemplar patch ${\textbf{\emph{x}}}_{i}$ of ${\textbf{\emph{R}}}_i \textbf{\emph{x}}$ and its similar patch ${\textbf{\emph{x}}}_{i,k}$, \ie, ${\textbf{\emph{w}}_{i,k} = \frac{1}{W}{\rm exp}(-\|{\textbf{\emph{x}}}_{i}- {\textbf{\emph{x}}}_{i,k}\|_2^2/h)}$, where $h$ and $W$ denote a predefined constant and a normalization factor \cite{Buades2005A}, respectively. It is worth noting that the reference matrix and the recovered matrix in the RRC model are both updated gradually and jointly in each iteration. More importantly, in the RRC approach, the rank minimization problem is solved by minimizing the rank residual rather than directly estimating LR matrices from corrupted observations as the same as in many classic rank minimization algorithms \cite{cai2010singular,Gu2014Weighted}.

\vspace{-0.15in}
\subsection{The Relationship between GSR and LR Models}
\label{sec:3.5}

Beyond exploiting different NSS priors by learning, recent research has investigated the relationship between GSR and LR models.

\vspace{-0.05in}
\subsubsection{LR-GSC}
\label{sec:3.5.1}
Recent work proposed the LR-GSC approach \cite{Zha2021Reconciliation} that simultaneously exploits sparsity and low-rankness of the dictionary-domain coefficients for each group of similar patches. LR-GSC employs the regularizer $\boldsymbol\Re_{\rm NL} (\textbf{\emph{x}}) = \boldsymbol\Re_{\rm LR-GSC} (\textbf{\emph{x}})$, which contains both GSR and LR parts, namely,
{\setlength\abovedisplayskip{6pt}
\setlength\belowdisplayskip{6pt}
\begin{equation}
{\boldsymbol\Re_{\rm LR-GSC} (\textbf{\emph{x}}) \triangleq  \argmin\limits_{\{\textbf{\emph{A}}_i\}, \{\textbf{\emph{B}}_i\}}\sum\nolimits_{i=1}^n \{  \frac{1}{2\mu}\|{\textbf{\emph{R}}}_i \textbf{\emph{x}} -{\textbf{\emph{D}}}_i{\textbf{\emph{A}}}_i\|_F^2 + \lambda\|{\textbf{\emph{A}}}_i\|_1  +\frac{1}{2\rho}\|{\textbf{\emph{A}}}_i-{\textbf{\emph{B}}}_i\|_F^2+ \tau\|{\textbf{\emph{B}}}_i\|_*\},}
\label{eq:11}
\end{equation}}
where $\| \cdot \|_1$ is applied for the sparsity penalty and the nuclear norm $\| \cdot \|_*$ is used for forming the low-rank penalty.  In the LR-GSC model, the dictionary ${\textbf{\emph{D}}}_i$ is learned from the corresponding group via PCA operation in each iteration. A low-rank approximation ${\textbf{\emph{B}}}_i$ is jointly estimated for each group sparse matrix ${\textbf{\emph{A}}}_i$. Similar to classical GSR models, the optimal sparse codes $ \{ \hat{\textbf{\emph{A}}}_i \}_{i=1}^n$ are used to reconstruct the latent clean image. With such  guidance, the LR-GSC approach integrates different LR and GSR learning formulations into a unified CS framework (P2) by reconciling nonlocal image modeling.
\vspace{-0.05in}
\subsubsection{Analyzing the Group Sparsity based on LR Methods}
\label{sec:3.5.2}
Recent work analyzes group sparsity based on the LR scheme \cite{Zha2020Benchmark}, which designs an adaptive dictionary for each group ${\textbf{\emph{X}}}_i$ to bridge the gap between GSR and LR models. Specially, the SVD is applied to the group matrix ${\textbf{\emph{X}}}_i\in{\mathbb R}^{b\times m}$, \ie,
{\setlength\abovedisplayskip{4pt}
\setlength\belowdisplayskip{4pt}
\begin{equation}
{{\textbf{\emph{X}}}_i= {\textbf{\emph{U}}}_i{\boldsymbol\Delta}_i{\textbf{\emph{V}}}_i^T=\sum\nolimits_{j=1}^{c} \delta_{i,j}{\textbf{\emph{u}}}_{i,j}{\textbf{\emph{v}}}_{i,j}^T,}
\label{eq:12}
\end{equation}}
where ${\boldsymbol\Delta}_i={\rm diag}(\delta_{i,1}, \delta_{i,2},\dots, \delta_{i,{c}})$ is a diagonal matrix, ${c} = {\rm min}(b,m)$,  and here ${\textbf{\emph{u}}}_{i,j}\in{\mathbb R}^{b\times 1}, {\textbf{\emph{v}}}_{i,j}\in{\mathbb R}^{m\times 1}$ represent the columns of ${\textbf{\emph{U}}}_i\in{\mathbb R}^{b\times c}$ and ${\textbf{\emph{V}}}_i\in{\mathbb R}^{m\times c}$, respectively. Following this, each dictionary atom $\textbf{\emph{d}}_{i,j}\in{\mathbb R}^{b\times m}$ of the adaptive dictionary $\textbf{\emph{D}}_i\in{\mathbb R}^{b\times{(c\times m)}}$ for each group $\textbf{\emph{X}}_i$ is defined as,
{\setlength\abovedisplayskip{4pt}
\setlength\belowdisplayskip{4pt}
\begin{equation}
{\textbf{\emph{d}}_{i,j}={\textbf{\emph{u}}}_{i,j}{\textbf{\emph{v}}}_{i,j}^T, \ \ \ j=1,2,\dots,{c}.}
\label{eq:13}
\end{equation}}
Finally, the learned adaptive dictionary $\textbf{\emph{D}}_i$ is as follows,
{\setlength\abovedisplayskip{4pt}
\setlength\belowdisplayskip{4pt}
\begin{equation}
{\textbf{\emph{D}}_i=[\textbf{\emph{d}}_{i,1},\textbf{\emph{d}}_{i,2},\dots,\textbf{\emph{d}}_{i,{c}}].}
\label{eq:14}
\end{equation}}%
Then, a GSR-based optimization problem can be derived to be equivalent to the corresponding LR problem with the learned group dictionary $\textbf{\emph{D}}_i$, so that the sparse coefficients of each group can be measured by estimating the singular values of each group.

\begin{figure*}[!t]
\vspace{-2mm}
\centering
\begin{minipage}[b]{1\linewidth}
{\includegraphics[width = 1\textwidth]{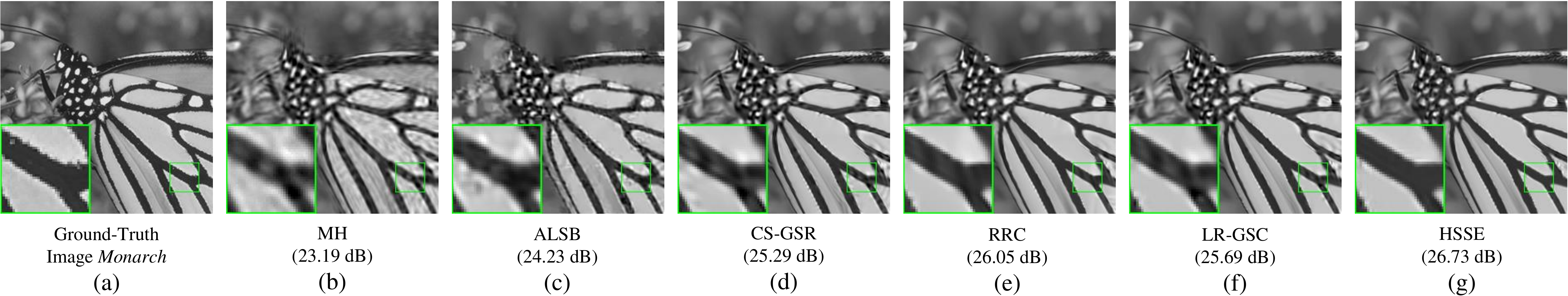}}
\end{minipage}
       \centering
\vspace{-8mm}
\caption{Illustration of image CS with a Gaussian random projection matrix at the block level (size of 32$\times$32) via different classes of algorithms: (a) Ground truth; (b) MH \cite{Chen2011Compressed}; (c) ALSB \cite{Zhang2014Image}; (d) CS-GSR \cite{Zhang2014Group}; (e) RRC \cite{Zha2020From}; (f) LR-GSC \cite{Zha2021Reconciliation}; and (g) HSSE  \cite{Zha2021Hybrid}.}
\label{fig:7}
\vspace{-8mm}
\end{figure*}

\vspace{-0.15in}
\subsection{The Pipeline of Nonlocal Structured Sparsity-based Image CS Algorithms}
\label{sec:3.6}
By embedding the learned regularizers~\eqref{eq:1.1},~\eqref{eq:3},~\eqref{eq:6},~\eqref{eq:8},~\eqref{eq:9}, and~\eqref{eq:11} into (P2), one can derive the image CS algorithms corresponding to CS-GSR \cite{Zhang2014Group},  GSRC \cite{Zha2020Group}, HSSE \cite{Zha2021Hybrid}, CS-NLR \cite{Dong2014Compressive}, RRC \cite{Zha2020From}, and LR-GSC \cite{Zha2021Reconciliation}, respectively. These algorithms usually need to solve several subproblems that involve efficient and closed-form solutions. While the actual forms of these regularizers (\ie,~\eqref{eq:1.1},~\eqref{eq:3},~\eqref{eq:6},~\eqref{eq:8},~\eqref{eq:9}, and~\eqref{eq:11}) vary, resulting in different nonlocal structured sparsity prior-based image CS algorithms, these algorithms all contain three main steps: 1) a block matching operation \cite{Dabov2007Image} is performed to form groups and each group contains many nonlocal similar patches; 2) sparse coding (update $\hat{\textbf{\emph{A}}}_i$) or rank minimization (update $\hat{\textbf{\emph{Z}}}_i$), and 3) update image $\hat{\textbf{\emph{x}}}$ in a least squares manner. Notably, the HSSE algorithm  \cite{Zha2021Hybrid} involves an additional external GSR learning step (\ie, ${\textbf{\emph{B}}}_i$ sub-problem in~\eqref{eq:6}) as it jointly imposes two complementary models on the image groups. In addition, the LR-GSC algorithm \cite{Zha2021Reconciliation} involves an additional LR estimation step (\ie, ${\textbf{\emph{B}}}_i$ sub-problem in~\eqref{eq:11}) as it investigates the LR property of group sparse coefficients in GSR.

\vspace{-0.15in}
\subsection{Empirical Comparisons}
\label{sec:3.7}

For a simple qualitative comparison of different classes of image CS algorithms, we obtain reconstructions of two classic images $\emph{Monarch}$ and $\emph{Fence}$, shown in Fig.~\ref{fig:7} and Fig.~\ref{fig:8}, respectively. For a fair comparison, CS measurements are sampled from the original image using a Gaussian random projection matrix at the block level (\ie, block-based CS \cite{Zhang2014Group}), and then image CS algorithms are applied to reconstruct the image using the sampled CS measurements.  
By using a 0.1 sampling rate for the two images, Fig.~\ref{fig:7} and Fig.~\ref{fig:8} show the reconstructions with peak signal-to-noise ratio (PSNR) values in decibels (dB), along with local image zoom-ins. In these empirical comparisons, we can observe that the learning-based methods consistently obtain higher PSNR values and visual results than the analytical transform method, \ie, MH \cite{Chen2011Compressed}. Compared to the patch-based sparse representation approach: ALSB \cite{Zhang2014Image}, CS-GSR \cite{Zhang2014Group} exploits each group as the basic unit for sparse representation to reconstruct images with finer details. Compared to the CS-GSR approach, the LR-GSC \cite{Zha2021Reconciliation} explores the LR property of group sparse coefficients in GSR and thus obtains better reconstruction results. Compared to single NSS prior-based methods, the HSSE scheme~\cite{Zha2021Hybrid}  jointly exploits internal and external NSS priors, which provide important complementary information for better reconstruction. Moreover, compared to all nonlocal structured sparsity prior-based methods, the deep learning method: CS-Net \cite{Shi2020Image} does not consider image NSS prior and thus cannot reconstruct well for images with  many similar or repeated textures. Some similar results have been obtained in more detailed quantitative comparisons in~\cite{Zha2020Group,Zha2021Hybrid,Zha2021Reconciliation}. The source code of these image CS methods is available at: \url{https://github.com/zhazhiyuan/Nonlocal_Image_CS_Demo.git}.

\begin{figure*}[!t]
\vspace{-2mm}
       \centering
\begin{minipage}[b]{1\linewidth}
{\includegraphics[width = 1\textwidth]{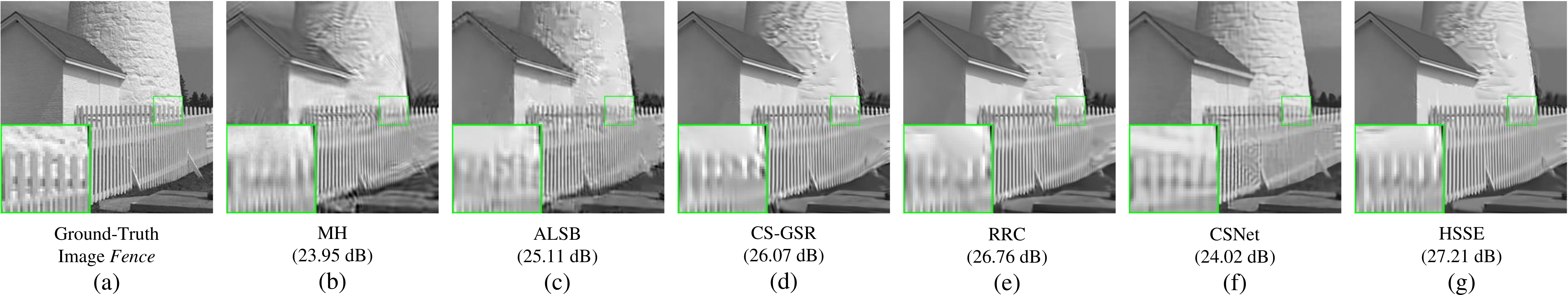}}
       \centering
\end{minipage}
\vspace{-8mm}
\caption{Illustration of image CS with a Gaussian random projection matrix at the block level (size of 32$\times$32) via different classes of algorithms: (a) Ground truth; (b) MH \cite{Chen2011Compressed}; (c) ALSB \cite{Zhang2014Image}; (d) CS-GSR \cite{Zhang2014Group}; (e) RRC \cite{Zha2020From}; (f) CS-Net \cite{Shi2020Image}; and (g) HSSE \cite{Zha2021Hybrid}.}
\label{fig:8}
\vspace{-8mm}
\end{figure*}

\vspace{-0.15in}
\section{Open Problems and Future Directions in Learning-based Image CS}
\label{sec:4}
Though learning-based methods have been extensively studied for image CS, there still exist several challenges and open problems. First, while many nonlocal structured sparsity prior-based reconstruction algorithms have been proposed, they usually enforce a uniform regularization parameter empirically for all image groups, which is unreasonable because they do not consider the differences across various groups. Although several approaches designed adaptive regularization parameters based on statistics, they cannot obtain promising reconstruction results for image CS. Developing a systematic regularization parameter selection method in nonlocal modeling with a principled way to obtain superior performance is an interesting line of research.

Second, most existing nonlocal structured sparsity-based models rely on matrix operations because they expand each image patch of the group into a vector and compose all vectors into a matrix. However, such schemes ignore the geometric structure of each local image patch and thus inevitably damage the intrinsic structural information of image patches. Since all nonlocal similar patches  (2D) can be
stacked to form a three-dimensional volume, it can be combined with structured sparsity prior to conduct nonlocal tensor modeling that effectively preserves the geometric structure of each image patch. Extending nonlocal image representation learning to effective  high-dimensional tensor modeling is an interesting line of research.

\begin{table*}[!t]
\vspace{-2mm}
\caption{Pros and cons of model-based optimization methods and deep neural network-based methods for image reconstruction.}
\vspace{-4mm}
\centering
\resizebox{0.65\textwidth}{!}				
{
\Huge
\centering
\begin{tabular}{|c|c|c|}
\hline
& \textbf{Model-based optimization methods} & \textbf{Deep neural networks-based methods}\\
\hline
\textbf{Pros} & \begin{tabular}[c]{c@{}c@{}}
\Checkmark Good generalization\\
\Checkmark Good interpretability
\end{tabular}
& \begin{tabular}[c]{c@{}c@{}}
\Checkmark Data-driven end-to-end learning\\
\Checkmark High-efficiency for superior reconstruction
\end{tabular}\\
\hline
\textbf{Cons} & \begin{tabular}[c]{c@{}c@{}}
\XSolidBold Weak handcrafted priors\\
\XSolidBold Heavy computational consumption\\ for optimization
\end{tabular}
&\begin{tabular}[c]{c@{}c@{}}
\XSolidBold Weak generalization\\
\XSolidBold Black box (limited interpretability)
\end{tabular}\\
\hline
\end{tabular}
}
\label{Tab:2}
\vspace{-8mm}
\end{table*}

Third, although both model-based optimization and supervised deep learning methods have been proposed for image CS, they have their own pros and cons, as shown in Table~\ref{Tab:2}. For example, supervised deep learning (neural network-based) methods learn image properties from training datasets with an end-to-end approach, which can be highly adaptive to the distribution of the training data, and thus obtain excellent reconstruction results. However, supervised deep learning methods often do not generalize well to new instances because the data for reconstructing a specific modality deviates from the distribution of the training corpus. In addition, the deep neural network is typically a black box, which cannot provide sound
interpretability. On the other hand, model-based optimization methods can capture quite general properties of image sets and generalize well. While model-based optimization approaches (\eg, GSR \cite{Zhang2014Group,Zha2021Hybrid,Zha2020Group} and LR \cite{Dong2014Compressive,Zha2020Benchmark,Zha2020From}) are usually self-supervised learning using handcrafted priors with good interpretability ability, they cannot provide sufficiently strong power for reconstruction and are usually time-consuming. Moreover, most existing model-based optimization and supervised deep learning methods have only considered internal or external image NSS priors, which largely ignored their important complementary properties. Developing a hybrid model that integrates the advantages of model-based optimization and supervised deep learning methods to avoid their respective drawbacks is an interesting line of research.

\vspace{-0.15in}
\section{Conclusions}
\label{sec:5}
This article briefly reviewed the timeline of image CS and discussed, in particular, some of the advances in nonlocal structured sparsity using GSR and LR modeling for image reconstruction. The nonlocal structured sparsity using GSR and/or LR modeling enables closed-form updates in the iterative algorithms, convergence, merging a variety of model properties, and effectiveness in reconstruction. We discussed a general learning-based regularization framework. In this context, we discussed a variety of GSR and LR learning regularizers based on their model structures and properties. Beyond employing different NSS priors by learning, we discussed the relationship between GSR and LR models. We discussed several existing challenges and ongoing directions in the field, such as a more rigorous understanding of the pros and cons of model-based optimization and supervised deep learning approaches in image reconstruction.

\section*{Acknowledgments}
This research is supported in part by National Natural Science Foundation of China under Grant U19A2052, 62020106011, 62271414, and 61971476, in part by the Ministry of Education, Republic of Singapore, under its Academic Research Fund Tier 1 (Project ID: RG61/22) and Start-up Grant, in part by Westlake Foundation (2021B1501-2) and the Research Center for Industries of the Future (RCIF) at Westlake University, in part by the Macau Science and Technology Development Fund, Macau SAR (File no. SKLIOTSC-2021-2023, 0022/2022/A1, 077/2018/A2, 0060/2019/A1, 0072/2020/AMJ). This work was carried out at the Rapid-Rich Object Search (ROSE) Lab, Nanyang Technological University, Singapore.

\section*{Authors}
\textbf{\emph{Zhiyuan Zha}} (zhiyuan.zha@ntu.edu.sg) received the Ph.D. degree with the School of Electronic Science and Engineering, Nanjing University, Nanjing, China, in 2018. He is a Research Fellow with the School of Electrical and Electronic Engineering, Nanyang Technological University, Singapore. His current research interests include inverse problems in image/video processing, sparse signal representation and machine learning. He was a recipient of the  Platinum Best Paper Award and  the Best Paper Runner Up Award at the IEEE International Conference on Multimedia and Expo  in 2017 and 2020, respectively.\\
\textbf{\emph{Bihan Wen}}  (bihan.wen@ntu.edu.sg) received the B.Eng degree in Electrical and Electronic Engineering from Nanyang Technological University, Singapore, in 2012, the M.S and Ph.D. degrees in Electrical and Computer Engineering from the University of Illinois at Urbana-Champaign, in 2015 and 2018, respectively. Currently, he is a Nanyang Assistant Professor with the School of Electrical and Electronic Engineering, Nanyang Technological University, Singapore. His current research interests include machine learning, computer vision, image and video processing, and computational imaging. Dr. Wen was a recipient of the 2016 Yee Fellowship and the 2012 Professional Engineers Board Gold Medal of Singapore and is currently a member of the IEEE Computational Imaging Technical Committee. He co-authored a paper that received the Top 10\% Best Paper Award at the IEEE International Conference on Image Processing in 2014, and another received the Best Paper Runner-Up Award at the IEEE International Conference on Multimedia and Expo in 2020.\\
\textbf{\emph{Xin Yuan}} (xyuan@westlake.edu.cn) is currently an Associate Professor at Westlake University in Hangzhou, Zhejiang, China. He was a video analysis and coding lead researcher at Bell Labs, Murray Hill, NJ USA from 2015 to 2021. Prior to this, he had been a Post-Doctoral Associate with the Department of Electrical and Computer Engineering, Duke University from 2012 to 2015, where he was working on compressive sensing and machine learning. Before joining Duke, Dr. Yuan obtained his B.Eng and M.Eng from Xidian University in 2007 and 2009, respectively, and his Ph.D. from the Hong Kong Polytechnic University in 2012. He was a recipient of  the Best Paper Runner Up Award at the IEEE International Conference on Multimedia and Expo  in  2020. He has been the Associate Editor of PATTERN RECOGNITION since 2019 and is leading the special issue of ``Deep Learning for High-Dimensional Sensing" in the IEEE JOURNAL OF SELECTED TOPICS IN SIGNAL PROCESSING in 2021.\\
\textbf{\emph{Saiprasad Ravishankar}} (ravisha3@msu.edu) is currently an Assistant Professor in the Departments of Computational Mathematics, Science and Engineering, and Biomedical Engineering at Michigan State University. He received the B.Tech. degree in Electrical Engineering from the Indian Institute of Technology Madras, India, in 2008, and the M.S. and Ph.D. degrees in Electrical and Computer Engineering in 2010 and 2014, respectively, from the University of Illinois at Urbana-Champaign, where he was then an Adjunct Lecturer and a Postdoctoral Research Associate. Since August 2015, he was a postdoc in the Department of Electrical Engineering and Computer Science at the University of Michigan, and then a Postdoc Research Associate in the Theoretical Division at Los Alamos National Laboratory from August 2018 to February 2019. His interests include signal and image processing, biomedical and computational imaging, machine learning, inverse problems, and large-scale data processing and optimization. He has received multiple awards including the Sri Ramasarma V Kolluri Memorial Prize from IIT Madras and the IEEE Signal Processing Society Young Author Best Paper Award for 2016 for his paper "Learning Sparsifying Transforms" published in the IEEE Transactions on Signal Processing. A paper he co-authored won a best student paper award at the IEEE International Symposium on Biomedical Imaging (ISBI) 2018, and other papers were award finalists at the IEEE International Workshop on Machine Learning for Signal Processing (MLSP) 2017 and ISBI 2020. He is currently a member of the IEEE Computational Imaging Technical Committee. He has organized several special sessions or workshops on computational imaging themes including at the Institute for Mathematics and its Applications (IMA), the IEEE Image, Video, and Multidimensional Signal Processing (IVMSP) Workshop 2016, MLSP 2017, ISBI 2018, and the International Conference on Computer Vision (ICCV) 2019 and 2021.\\
\textbf{\emph{Jiantao Zhou}} (jtzhou@um.edu.mo) received the B.Eng degree from the Department of Electronic Engineering, Dalian University of Technology, in 2002, the M.Eng degree from the Department of Radio Engineering, Southeast University, in 2005, and the Ph.D. degree from the Department of Electronic and Computer Engineering, Hong Kong University of Science and Technology, in 2009. He held various research positions with the University of Illinois at UrbanaC-Champaign, the Hong Kong University of Science and Technology, and the McMaster University. He is currently an Associate Professor with the Department of Computer and Information Science, Faculty of Science and Technology, University of Macau. He holds four granted U.S. patents and two granted Chinese patents. His research interests include multimedia security and forensics, and multimedia signal processing. He has co-authored three papers that received the Best Paper Award at the IEEE Pacific-Rim Conference on Multimedia in 2007, the Best Student Paper Award and the Best Paper Runner Up Award at the IEEE International Conference on Multimedia and Expo in 2016 and 2020, respectively.
He has been the Associate Editor of IEEE TRANSACTIONS ON IMAGE PROCESSING since 2018 and the Associate Editor of IEEE TRANSACTIONS ON MULTIMEDIA since 2021.\\
\textbf{\emph{Ce Zhu}} (eczhu@uestc.edu.cn) received the B.S. degree from Sichuan University, Chengdu, China, in 1989, and the M.Eng and Ph.D. degrees from Southeast University, Nanjing, China, in 1992 and 1994, respectively, all in electronic and information engineering. He held a post-doctoral research position with the Chinese University of Hong Kong in 1995, the City University of Hong Kong, and the University of Melbourne, Australia, from 1996 to 1998. He was with Nanyang Technological University, Singapore, for 14 years from 1998 to 2012, where he was a Research Fellow, a Program Manager, an Assistant Professor, and then promoted to an Associate Professor in 2005. He has been with University of Electronic Science and Technology of China, Chengdu, China, as a Professor since 2012. His research interests include video coding and communications, video analysis and processing, 3D video, visual perception and applications. He has served on the editorial boards of a few journals, including as an Associate Editor of IEEE TRANSACTIONS ON IMAGE PROCESSING, IEEE TRANSACTIONS ON CIRCUITS AND SYSTEMS FOR VIDEO TECHNOLOGY, IEEE TRANSACTIONS ON BROADCASTING, IEEE SIGNAL PROCESSING LETTERS, an Editor of IEEE COMMUNICATIONS SURVEYS AND TUTORIALS, and an Area Editor of SIGNAL PROCESSING: IMAGE COMMUNICATION. He has also served as a Guest Editor of a few special issues in international journals, including as a Guest Editor in the IEEE JOURNAL OF SELECTED TOPICS IN SIGNAL PROCESSING. He is an APSIPA Distinguished Lecturer (2021-2022), and was also an IEEE Distinguished Lecturer of Circuits and Systems Society (2019-2020). He is a co-recipient of multiple paper awards at international conferences, including the most recent Best Demo Award in IEEE MMSP 2022, and the Best Paper Runner Up Award in IEEE ICME 2020.\\

\vspace{-0.15in}
{\footnotesize
\bibliographystyle{IEEEtran}
\bibliography{ngslr_ref}
}

\end{document}